\documentclass[preprint]{aastex63}

\usepackage[utf8]{inputenc}
\usepackage{natbib}
\usepackage{color}
\usepackage{graphicx}
\usepackage{rotating}
\usepackage{hyperref}
\usepackage[all]{hypcap}
\usepackage{mathrsfs}
\usepackage{amsmath}
\usepackage{amssymb}
\usepackage{float}
\usepackage{graphicx}
\usepackage{gensymb}
\usepackage{subfigure}
\usepackage{lineno}
\hypersetup{
    unicode=false,          
    pdftoolbar=true,        
    pdfmenubar=true,        
    pdffitwindow=false,     
    pdfstartview={FitH},    
    pdftitle={My title},    
    pdfauthor={Author},     
    pdfsubject={Subject},   
    pdfcreator={Creator},   
    pdfproducer={Producer}, 
    pdfkeywords={keyword1} {key2} {key3}, 
    pdfnewwindow=true,      
    colorlinks=true,        
    linkcolor=blue,         
    citecolor=blue,         
    filecolor=magenta,      
    urlcolor=cyan           
}

\newcommand{\p}[1]{{\color{red}{#1}}}

\begin{document}
\title{Data-constrained Magnetohydrodynamic Simulation of a 
Filament Eruption in a Decaying Active Region 13079 on a Global Scale}

\author[0009-0003-0312-2513]{Yihua Li}
\affil{School of Astronomy and Space Science and Key Laboratory of Modern Astronomy and Astrophysics, Nanjing University, Nanjing 210023, China}

\author[0000-0002-9293-8439]{Yang Guo}
\affil{School of Astronomy and Space Science and Key Laboratory of Modern Astronomy and Astrophysics, Nanjing University, Nanjing 210023, China} \email{guoyang@nju.edu.cn}

\author{Jinhan Guo}
\affil{School of Astronomy and Space Science and Key Laboratory of Modern Astronomy and Astrophysics, Nanjing University, Nanjing 210023, China} 

\author[0000-0002-4978-4972]{M. D. Ding}
\affil{School of Astronomy and Space Science and Key Laboratory of Modern Astronomy and Astrophysics, Nanjing University, Nanjing 210023, China} 

\author[0000-0002-7153-4304]{Chun Xia}
\affil{School of Physics and Astronomy, Yunnan University, Kunming 650050, China}

\author[0000-0003-3544-2733]{Rony Keppens}
\affil{Centre for mathematical Plasma Astrophysics, Department of Mathematics, KU Leuven, Celestijnenlaan 200B, B-3001 Leuven, Belgium}

\begin{abstract}
Filaments are special plasma phenomena embedded in the solar atmosphere, characterized by unique thermodynamic properties and magnetic structures. Magnetohydrodynamic (MHD) simulations are useful to investigate the eruption mechanisms of filaments. We conduct a data-constrained zero-$\beta$ MHD simulation in spherical coordinates to investigate a C3.5 class flare triggered by an eruptive filament on 2022 August 15 in a decaying weak active region 13079. We reconstruct the three-dimensional coronal magnetic field using vector magnetograms and synoptic maps from the Solar Dynamics Observatory/Helioseismic and Magnetic Imager (SDO/HMI). We transform vector magnetic field into Stonyhurst heliographic spherical coordinates combined with a synoptic map and constructed a potential field source surface (PFSS) model with a magnetic flux rope (MFR) embedded using the Regularized Biot--Savart Laws (RBSL). Subsequently, we conduct a spherical zero-$\beta$ MHD simulation using the Message Passing Interface Adaptive Mesh Refinement Versatile Advection Code (MPI-AMRVAC) and replicated the entire dynamic process of the filament eruption consistent with observations. With the calculation of time-distance profile, Qusai-Separatrix Layers (QSL), and synthetic radiation from simulated current density, we find a good agreement between our simulation and observations in terms of dynamics and magnetic topology. Technically, we provide a useful method of advanced data-constrained simulation of weak active regions in spherical coordinates. Scientifically, the model allows us to quantitatively describe and diagnose the entire process of filament eruption.  

\end{abstract}

\section{Introduction} 

Solar filaments are one of the most special phenomena embedded in the solar atmosphere. They exhibit unique structures and thermodynamic properties. To balance the gravity of dense plasma material, magnetic structures of filaments are required \citep{2007wangformation_filament}. Two common types of filament magnetic field channels are twisted magnetic flux ropes (MFRs) and sheared magnetic arcades \citep{Ouyang2017}. Moreover, filaments are not always stable, and their eruptions are related to various solar activities including flares and coronal mass ejections \citep[CMEs;][]{chenCoronalMassEjections2011,mccauleyProminenceFilamentEruptions2015, cheng2017origin}. According to the classification of filament eruptions in \citet{gilbert2007filament}, a successful or partial filament eruption can lead to CMEs and flare events, ejecting energy and plasma into interplanetary space, and modulating the space weather.

Magnetohydrodynamic (MHD) numerical simulations are effective to study the formation, evolution, triggering, and eruption mechanisms of filaments, which can help to precisely diagnose the coronal magnetic field structure, study the dynamics, topology and thermodynamic properties of filaments, and forecast space weather \citep{Chen2020}. Many studies revealed two main mechanisms in forming filament channels by numerical simulations, including direct emergence of flux ropes from the convection zone \citep{fan2001emergence, okamoto2008emergence}, or the formation through magnetic reconnection between shear arcades \citep{van1989formation}. \citet{patsourakos2020decoding} classified the formation mechanisms of MFRs into three categories: flux emergence, flux cancellation and helicity condensation. These mechanisms might interact with each other and form MFRs through joint effects. For the filament material, it is believed that there is insufficient plasma in the corona to condense into the filaments. Therefore, plasma material is transported from the chromosphere \citep{2017songfilament}, which can be achieved through mechanisms including injection from footpoints \citep{an1985formation}, lifting up with magnetic field evolution \citep{okamoto2008emergence,zhao2017formation}, and evaporation-condensation process \citep{1991antiochosfilament}. Among them, the evaporation-condensation model has been extensively verified by numerical simulations \citep{2016xia_ronyfilament,zhou2020simulations,guo2021formation}. For eruption mechanisms, \citet{2021NatCoZhongze} reproduced a confined eruption with MHD simulations and found that the Lorentz force components from the non-axisymmetry of the MFR constrained the eruption. Additionally, radiative magnetohydrodynamic (rMHD) simulations, which utilize more sophisticated energy equations, facilitate a direct comparison between model-generated observables and actual observational data. \citet{2022ApJChenfengRMHDflux} and \citet{2022ApJWangcanRMHDplasma,2023ApJWangcanRMHDforces} conducted detailed rMHD simulations of flux emergence in an active-region scale from the convection zone to the corona and analyzed plasma thermodynamics, magnetic topology, and force evolution of the confined MFR eruption.

We use data-driven or data-constrained MHD simulations, where observed magnetic field is used as the initial and/or boundary conditions, to study the evolution of material density, velocity, and magnetic topology of filaments. Since coronal magnetic field measurements are difficult, we often use indirect methods such as Non-Linear Force Free Field (NLFFF) extrapolation \citep{low1990modeling, titov1999basic, guo2016magneto1, guo2016magneto2} to search for MFRs. However, NLFFF techniques usually fail to produce MFRs for filaments in decaying or quiescent regions, whose magnetic field is relatively weak and MFRs sometimes detach from the photosphere \citep{guo2023data}. Therefore, constructing and embedding MFRs using numerical methods such as arcade flux rope insertion \citep{malanushenko2014using, dalmasse2019data} and Regularized Biot–Savart Laws \citep[RBSL;][]{titov2014method, titov2018regularized}, are more flexible and reliable. These methods can be combined with multi-perspective observations and three-dimensional path reconstruction \citep{torokWritheHelicalStructures2010,guo2019mfr,xuThreedimensionalReconstructionFilament2020} with detailed information of the radius, axis path, magnetic flux, and electric current of the embedded MFR.

Aforementioned techniques are suitable for filaments in decaying active regions. \citet{guo2023data} have implemented a RBSL method in spherical coordinates in the Message Passing Interface Adaptive Mesh Refinement Versatile Advection Code \citep[MPI-AMRVAC;][]{2003ronyamrvac,porth2014mpi,xia2018mpi,keppens2021mpi,2023ronyamrvac3.0}. Based on the RBSL method and Potential Field Source Surface (PFSS) model, we conduct a data-constrained zero-$\beta$ MHD simulation in the spherical coordinate system to study a filament eruption event in active region 13079. Section~\ref{sec:observation} describes the observation of this event. Section~\ref{sec:magnetic_field_reconstruction} presents the magnetic field reconstruction process using numerical methods. Our data-constrained MHD simulation and result analysis are displayed in Section~\ref{sec:mhd simulation}. A summary and discussions are given in Section~\ref{sec:summary}.

\section{Observation} \label{sec:observation}

In the present investigation, we concentrate on a C3.5 flare that commenced at 04:25 UT on 2022 August 15, in NOAA active region 13079. This event is special for two main reasons. First, the position of 13079 was close to the solar limb as observed by the Solar Dynamics Observatory (SDO) when this flare happened.  It occupied a large area, where the curvature of the region cannot be ignored. Consequently, using conventional simulation methods in Cartesian coordinates may introduce larger computation errors, which requires to use numerical methods in spherical coordinates. Secondly, this flare was triggered by a filament, distinctly visible in multiple wavebands from the Atmospheric Imaging Assembly (AIA) on board SDO. This facilitates the feasibility of embedding magnetic flux ropes (MFRs) and conducting MHD simulations in spherical coordinates.

We primarily use magnetic field data from the Helioseismic and Magnetic Imager \citep[HMI;][]{2012SoPhsdohmi, 2012SoPhsdohmiground}) on board SDO. The cadence of line-of-sight full-disk magnetograms is $45$ s, and it is $720$ s for vector magnetograms, and the pixel size is $0.5^{''}$. SDO/AIA provides multiple extreme-ultraviolet (EUV) wavebands of full-disk coronal images with a pixel size of $0^{''}.6$, a temporal cadence of 12 s and a field of view of $1.3 R_\sun$.

\begin{figure*}
    \centering
    \includegraphics[width=0.7\textwidth]{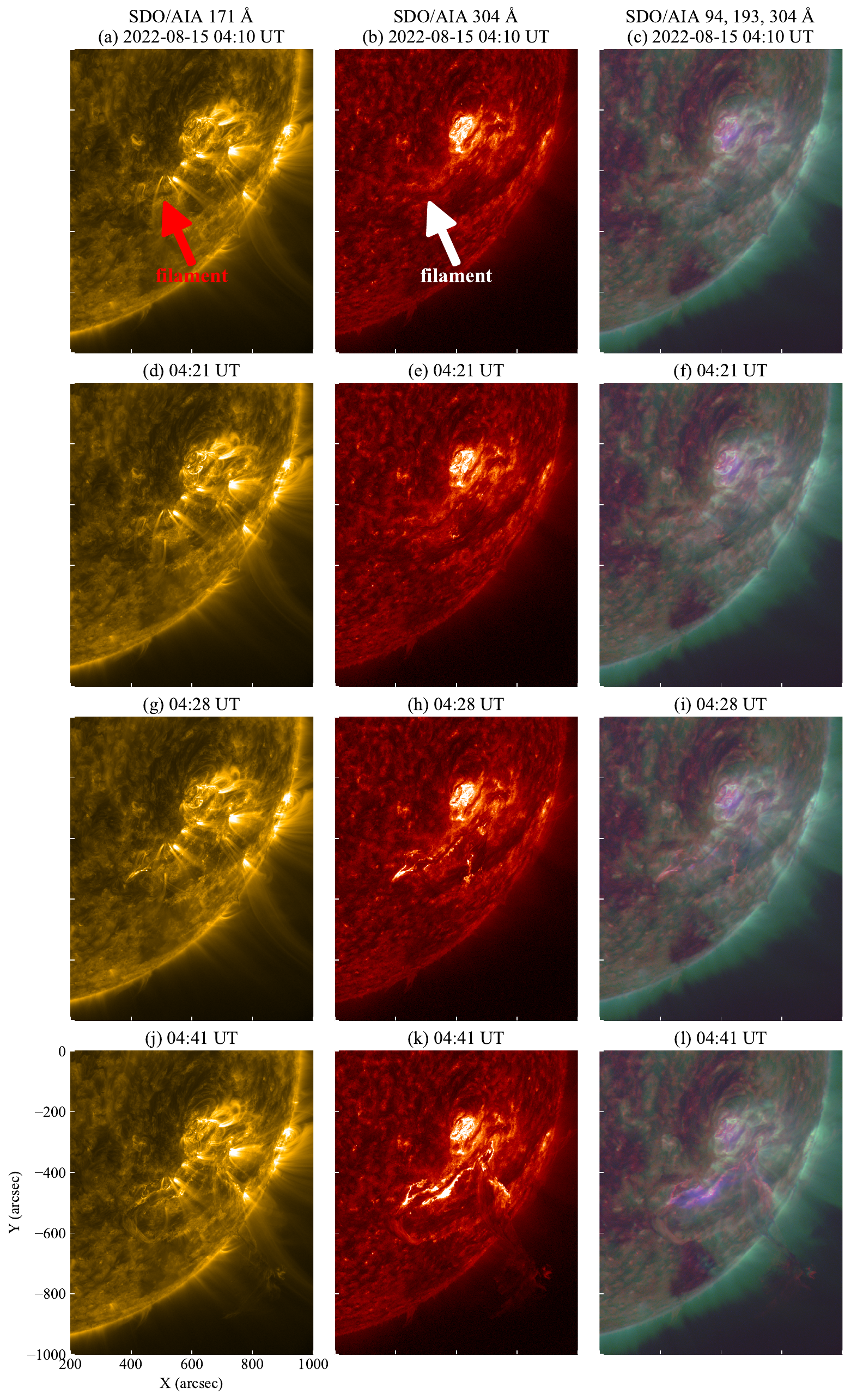}
    \caption{Evolution of the filament in active region 13079 on 2022 August 15 in multiple SDO/AIA wavelengths. (a, d, g, j) 171 \AA \ images at 04:10, 04:21, 04:28, and 04:41 UT respectively. (b, e, h, k) 304 \AA \ images. (c, f, i, l) Composite images constructed by 94, 193, and 304 \AA \ observations in the red, green, and blue channels, respectively. An animation showing the evolution of the observation from 04:00 to 04:59 UT is available online. }
    \label{fig:aia_observation}
\end{figure*}

The filament was suspended above a polarity inversion line (PIL) of the magnetic field from 03:30 UT. Its limbs lightened at around 04:16 UT, after which the filament began to lift up radially with brightening flare ribbons. A CME starting at 05:00 UT is observed in Large Angle and Spectrometric Coronagraph (LASCO) C2 on board Solar and Heliospheric Observatory (SOHO). Therefore, this is a successful filament eruption. We present the evolution process of the filament in 94 \AA, 171 \AA, 193 \AA \ and 304 \AA \ in Figures \ref{fig:aia_observation}.  The 171 \AA \ band corresponds to a coronal temperature at $\sim 0.63$ MK, displaying static coronal structures including coronal loops at 04:10, 04:21, 04:28, and 04:41 UT (Figures \ref{fig:aia_observation}a, \ref{fig:aia_observation}d, \ref{fig:aia_observation}g, and \ref{fig:aia_observation}j). The 304 \AA \ band corresponds to $\sim$ 0.05 MK, revealing low-temperature structures from chromosphere and the transition region (Figures \ref{fig:aia_observation}b, \ref{fig:aia_observation}e, \ref{fig:aia_observation}h, and \ref{fig:aia_observation}k). Additionally, we display a three-color composite image synthesized from 94 \AA \ ($\sim$ 6.3 MK), 193 \AA \ ($\sim$ 1.25 MK) and 304 \AA, where the eruptive filament is visible in these three wavebands (Figures \ref{fig:aia_observation}c, \ref{fig:aia_observation}f, \ref{fig:aia_observation}i, and \ref{fig:aia_observation}l). The three-color images reveal multi-temperature structures. 

We summarize observational characteristics of this filament during the eruption as follows. Before the flare happened, brightening appeared within the active region. Subsequently, it gradually uplifted and plasma was ejected. The filament that triggered the flare is suspended above the PIL. Its left footpoint is located in a decaying and weak positive magnetic field region, while the right footpoint resides in a more active and strong negative field region. The spine of the filament is nearly parallel to the PIL, and the filament exhibits positive helicity. During the eruption of this filament, its uplift direction is almost radial without significant deviations towards the east or west side.

\section{Magnetic Field Reconstruction in Spherical Coordinates}
\label{sec:magnetic_field_reconstruction}

We first employ the RBSL model in MPI-AMRVAC to embed an MFR in the filament channel based on the measured filament axis and magnetic flux from observations. Then, we use MPI-AMRVAC to construct a PFSS model with a SDO/HMI synoptic map of Carrington rotation 2260. The synoptic map is combined with an instantaneous vector magnetic field in active region 13079, which makes it a synoptic frame. The magnetic field generated by the RBSL flux rope on the bottom boundary is also subtracted from the synoptic frame (Figure \ref{fig:rbsl_pfss}a). The PFSS field is finally computed by such a RBSL field subtracted synoptic frame. When we add the PFSS field and the RBSL field together, the radial magnetic field of the synoptic frame is restored. Finally, we use the magneto-frictional method to relax the PFSS+RBSL field to a nearly NLFFF state. In doing so, we make use of MPI-AMRVAC \citep{porth2014mpi, xia2018mpi, keppens2023mpi}, the Solar SoftWare (SSW) package in IDL, and the Magnetic Modeling Codes \citep[MMC\footnote{https://github.com/njuguoyang/magnetic\_modeling\_codes};][]{guoOriginStructuresSolar2017}.

\subsection{Observational Data Pre-processing}
\label{sec:data}

We use magnetic field data from SDO/HMI, including a vector magnetogram and a synoptic map, as boundary conditions for magnetic field reconstruction. Multi-step preprocessings are performed as follows. We select data series "$hmi.B.720s$", the vector magnetogram of active region 13079 at 04:00 UT on 2022 August 15. Sequential processing steps include removing the 180$^\circ$ ambiguity of the transverse components of the vector magnetogram, correcting projection effects using the rotation matrix in \citet{guoOriginStructuresSolar2017}, and eliminating Lorentz forces and torques on the photosphere \citep{wiegelmann2006preprocessing}. The projection correction converts the vector field firstly to a local Cartesian coordinates and then to the StonyHurst Heliographic coordinates, transforming ($B_x,~B_y,~B_z$) to ($B_r,~B_\theta,~B_\phi$). We note that $\theta$ is the complementary angle of the latitude, and $\phi$ is the longitude measured from the central meridian on the backside of the Sun. The detailed transformation method is described in \citet{guo2023data}.

\subsection{Regularized Biot--Savart Laws}\label{sec:rbsl}

\citet{titov2018regularized} proposed a new method to construct an MFR using regularized Biot--Savart laws (RBSL). The embedded MFR is in a force-free state, possessing an axis of arbitrary shape and different current distribution profiles in a circular cross-section. It has four adjustable parameters, including the minor radius $a$, the three-dimensional axis path $\mathcal{C}$, magnetic flux $F$, and electric current $I$. It was originally proposed in Cartesian coordinates and has been implemented in the spherical coordinate system \citep{guo2023data}.

The minor radius $a$ can be constrained with measurements of the filament width in SDO/AIA observations. More accurately, $a$ is usually 1 to 3 times larger than the filament width \citep{guo2019solar, kang2023modelling}. \citet{guoProminenceFineStructures2022a} indicates that the filament material occupies the lower quarter of the radial cross-section of the MFR from a theoretical estimate. Considering aforementioned relations, we measure the filament width from observations of SDO/AIA in 193 \AA, 211 \AA, and 304 \AA \ wavebands. It is approximately (1.75--2.10)$\times 10^9$ cm. Within a reasonable range, multiple values of $a$ are tested and we choose $a=2.0\times 10^9~\mathrm{cm}$ with the consideration that positions of footpoints of the embedded MFR should be consistent with observations. 

For the axis path $\mathcal{C}$, we use a parameter $R$, the major radius of the MFR, to limit the maximum height of it. With this major radius, we can control the morphology of the main axis curve with simple geometric methods described in details in \citet{xuThreedimensionalReconstructionFilament2020}. The path is calculated with the same method as \citet{xuThreedimensionalReconstructionFilament2020} with Equations (5)--(8) and $\theta=0$. We select different major radius $R$ to construct the three-dimensional axis path of the MFR, repeat the measurements by trial-and-error, and compare the constructed MFR with the observed filament to determine the best matched path. Eventually, we set $R=90$ Mm for the following reasons. First, the observed filament is in a weak and decaying active region, keeping stable for days before eruption. The major radius of the filament may approach the empirical height of intermediate filaments. Secondly, we find that constructed MFRs with major radii lower than 70 Mm poorly match the observed filament, and even show a tendency to sink below the photosphere during MHD simulations. The selected path projected on photospheric radial magnetic field and the SDO/AIA 304 \AA \ image is shown in Figures \ref{fig:rbsl_pfss}b and \ref{fig:rbsl_pfss}c. Note that the path is a projection of the three-dimensional filament on the photosphere, so it does not entirely match with the observed filament spine because of this projection as shown in Figure \ref{fig:rbsl_pfss}c.

For the magnetic flux $F$ of the MFR, we over-plot the two footpoints on the radial magnetic field  as shown in Figure \ref{fig:rbsl_pfss}b and derive fluxes at two footpoints, namely $F_1$ and $F_2$. Then, we calculate the average flux $F_0$ of the absolute values of $F_1$ and $F_2$, and find that $F_0=(|F_1|+|F_2|)/2=5.03\times 10^{20}$ Mx. The flux of the embedded MFR is varied with different values as follows: $F=F_0,~2F_0,~4F_0,~8F_0,~10F_0,~12F_0,~20F_0$. Considering the consistency of MHD simulations with observations, we choose $F=12F_0$ as the optimal flux. As for lower fluxes including $F=F_0,~2F_0,~4F_0,~8F_0,~10F_0$, the embedded MFR would sink below the photosphere surface with a low major radius $R$ or simply unwind with a higher $R$, indicating am inconsistency with the observed filament eruption. For higher fluxes including $20F_0$ and etc., the twist of the MFR is excessively large and, due to kink instability, significant entanglement arises in MHD simulations, which is also unreasonable. For the electric current $I$ of the MFR is computed as $F=\pm 3/(5\sqrt{2})\mu I a 
$ \citep{titov2018regularized} to satisfy the internal equilibrium of the MFR, and a positive sign is selected because of the filament exhibits positive magnetic helicity referring to methods in \citet{chen2014imaging} and \citet{ouyang2015flux}. The results of the magnetic field reconstruction is shown in Figure \ref{fig:rbsl_pfss}d with the bottom boundary displaying the radial magnetic field on $r=1.01R_\odot$ plane of the PFSS+RBSL model. The footpoints of the MFR exhibit relatively high magnetic field intensity because it shows coronal magnetic field instead of that on photosphere.

This completely determines the initial MFR using the RBSL method. We now detail how to initialize the surrounding field by PFSS (Section \ref{sec:pfss}), and the total configuration is then relaxed to an NLFFF configuration (Section \ref{sec:mf}).

\subsection{Potential Field Source Surface Model}
\label{sec:pfss}

PFSS model is generally used to describe the large-scale magnetic field of the solar corona in the spherical coordinate system \citep{1969SoPh....6..442S, 1992ApJ...392..310W}. It provides a simple but efficient approximation of global magnetic field. We assume that small current sheets in high corona would not significantly affect the global magnetic structure and the magnetic field is in a force-free and current-free state. 

\begin{figure*}
    \centering
    \includegraphics[width=0.8\textwidth]{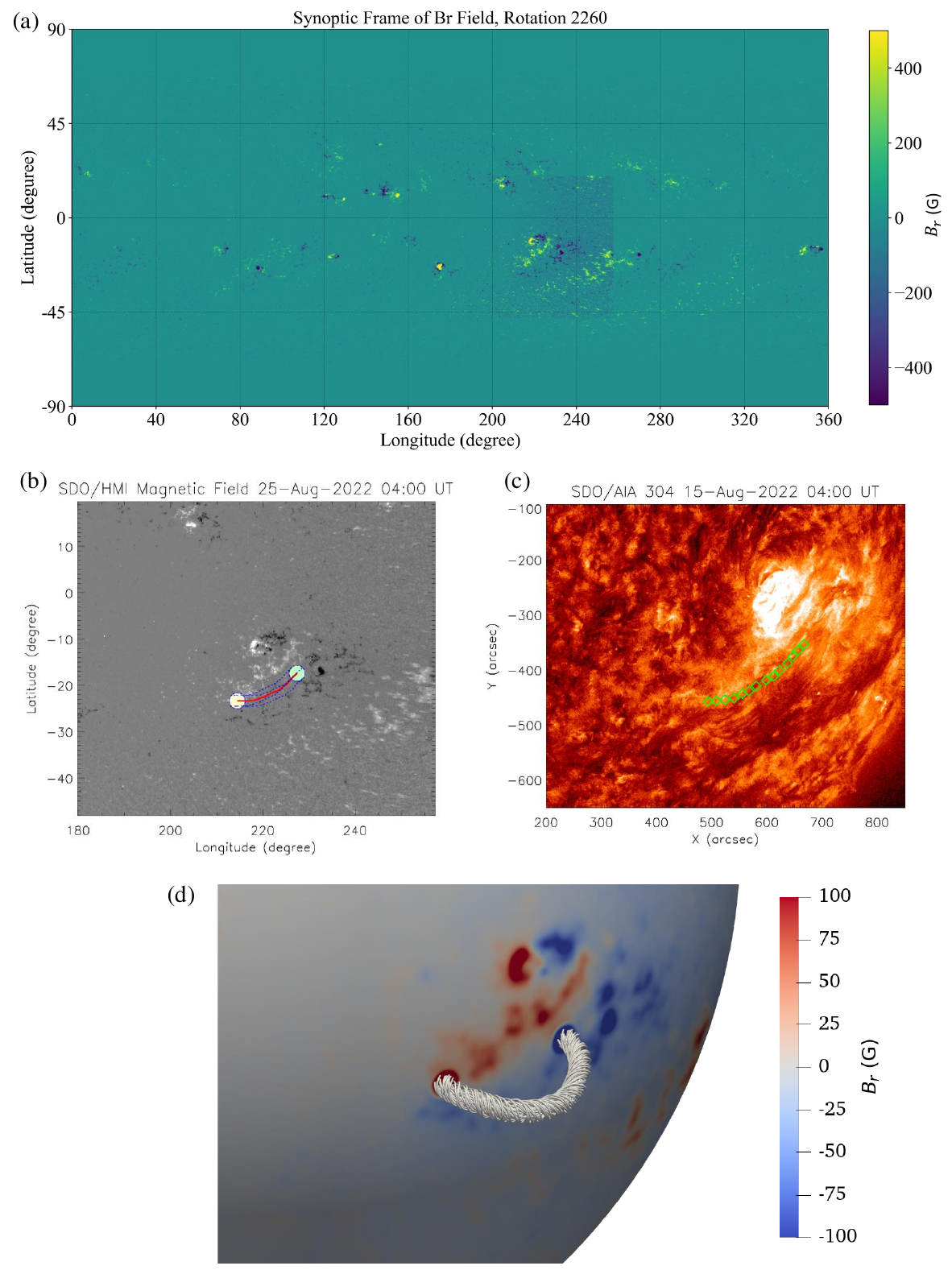}
    \caption{Synoptic frame, filament axis path overlaid on magnetic field and 304 \AA \ image, and the reconstructed coronal magnetic field. (a) $B_r$ of SDO/HMI vector magnetic field map embedded in the SDO/HMI synoptic map of Carrington rotation 2260, serving as the boundary condition for the PFSS model. (b) Projected filament axis path on SDO/HMI $B_r$ at 04:00 UT. Green and yellow circles denote the two footpoints of the MFR. The red and blue lines show the projected path. (c) Filament axis path overlaid on the SDO/AIA 304 \AA \ image at 04:00 UT. Green diamonds represent the projected filament path on the solar surface. (d) The embedded MFR constructed by the RBSL method, with the bottom boundary displaying $B_r$ on $r=1.01R_\odot$ plane.}
    \label{fig:rbsl_pfss}
\end{figure*}

We use two boundary conditions for the PFSS model, one is the radial magnetic field data on the solar surface $r=R_\sun$, and the other one is an assumption that the magnetic field is purely radial at the source surface $r=2.5R_\sun$. To construct the boundary on the solar surface, we use a synoptic map of Carrington rotation 2260 from SDO/HMI. We note that the synoptic map is calculated based on line-of-sight (LOS) photospheric magnetograms along the central meridian over 27 days, which is approximately one rotation period. However, at the time of the C3.5 flare, active region 13079 already moved close to the right limb of the solar disk with non-negligible changes of photospheric magnetogram. Therefore, using the synoptic map as boundary condition is inadequate. We incorporate the preprocessed vector magnetic field of active region 13079 into the synoptic map at corresponding positions to obtain the radial magnetic field boundary that simultaneously contains global information and the local high resolution map of the active region. The combined radial magnetic field is called a synoptic frame shown in Figure \ref{fig:rbsl_pfss}a. We use it as the inner boundary condition at $r=1R_\sun$ for the PFSS model.

Moreover, according to the design of the RBSL method, $B_z$ only has values at two circular footpoints of the MFR. If the two footpoints are not too far apart, the radial magnetic field $B_r$ at photosphere of the MFR will also be concentrated at footpoints in spherical coordinates \citep{guo2023data}. To ensure the accuracy of the photospheric magnetic field and prepare the final boundary conditions for the PFSS model, it is necessary to remove the strong radial magnetic field at the footpoints of the inserted flux rope. After aforementioned preparations, we calculate the PFSS model in MPI-AMRVAC with the module that was first demonstrated in \citet{porth2014mpi}. The next step is to relax the overall magnetic field (PFSS+RBSL) to an NLFFF field, using the magneto-frictional technique.

\subsection{NLFFF Relaxation Using Magneto-frictional Method}
\label{sec:mf}

After reconstructing the initial coronal magnetic field with the RBSL and PFSS models, the obtained magnetic field is still insufficient for MHD simulations. First, the large-scale magnetic field constructed by the PFSS model only utilizes the radial magnetic field $B_r$, while in MHD simulations, $B_\theta$ and $B_\phi$ are also required as boundary conditions. Secondly, the embedded MFR can spontaneously maintain internal equilibrium, but with the background field, it generally does not satisfy an overall external equilibrium. 

Therefore, we need to relax the combined magnetic field to a nearly NLFFF state, which is often used to describe the coronal magnetic field in active regions. We use the magneto-frictional method \citep{guo2016magneto1,guo2016magneto2} in MPI-AMRVAC to relax the NLFFF field. The bottom boundary is the vector magnetic field transformed to the StonyHurst Heliographic coordinates $B_r, ~B_\theta$ and $B_\phi$ at $r=1R_\sun$. We set the two factors $c_c$ and $c_y$ to 0.3 and 0.5, respectively, which control the relaxation speed of the magneto-frictional method defined in \citet{guo2016magneto1}. Additionally, the factor $c_d$, which controls speed to clean the divergence, is set to 0.01. After $3\times10^5$ steps of iteration, the force-free and divergence-free metrics, $\sigma_J$ and $\left \langle|f_i|\right\rangle$, defined in \citet{wheatland2000optimization}, are 0.343 and $1.10\times 10^{-5}$. These two parameters  are comparable to other previous models. \citet{guo2016magneto2} showed that after $10^5$ steps of iterations, $\sigma_J$ fell within the range of $[0.2,~0.5]$ for multiple events, and $\left \langle|f_i|\right\rangle$ was within $[1.0\times 10^{-5},~ 9\times 10^{-4}]$. Empirically, the magnetic field configurations approximate the force-free field relatively well with similar or smaller metrics, and are suitable for MHD simulations \citep{guo2023thermodynamic}. Compared with the original embedded MFR in Figure \ref{fig:rbsl_pfss}d, the twist of the relaxed MFR is significantly reduced, which can be seen in Figure \ref{fig:mhd_simulation1}a, and is more consistent with the observed filament.  

\section{Zero-$\beta$ MHD Simulations}
\label{sec:mhd simulation}
\subsection{Modeling Setup}

The plasma $\beta$, which is the ratio between the gas pressure to the magnetic pressure, is defined as $\beta=p/(B^2/2\mu_0)$. The solar corona in active regions can be approximated as in a low-$\beta$ condition. Therefore, a zero-$\beta$ MHD model is adopted to simulate the dynamics and topology of the coronal magnetic field. The zero-$\beta$ MHD model omits the gas pressure, gravity, and the energy equation. There is only the Lorentz force to change the momentum. The zero-$\beta$ MHD equations are expressed in a dimensionless form as follows:
\begin{gather}
    \label{mhd_rho}
    \frac{\partial \rho}{\partial t}+\nabla \cdot(\rho\mathbf{v})=0,\\
    \label{mhd_v}
    \frac{\partial \rho\mathbf{v}}{\partial t}+\nabla\cdot(\rho\mathbf{vv}-\mathbf{BB}+\frac{\mathbf{B}^2}{2}\mathcal{I})=0,\\\frac{\partial\mathbf{B}}{\partial t}+\nabla\cdot(\mathbf{vB}-\mathbf{Bv})=-\nabla\times (\eta\mathbf{J}),\\
    \mathbf{J}=\nabla\times\mathbf{B},
\end{gather}
where $\rho, ~\mathbf{v}$ and $\mathbf{B}$ are density, speed, and magnetic field, $\eta$ is the resistivity, and $\mathbf{J}$ is the electric current density. The computation domain is $\displaystyle [r_{min},r_{max}]\times[\theta_{min},\theta_{max}]\times[\phi_{min},\phi_{max}]=[1.001R_\odot,1.801R_\odot]\times[71.08\degree,137.58\degree]\times[180.57\degree, 256.81\degree]$. Note that since we transform the vector magnetic field to the StonyHurst Heliographic coordinates, $\theta$ is the complementary angle of the latitude and $\phi$ is the longitude measured from the central meridian on the back side of the Sun. The domain is resolved by $400\times280\times320$ cells, where the grids along $r$-, $\theta$-, and $\phi$-directions are uniform. The initial magnetic field is the RBSL+PFSS model after NLFFF relaxation using the magneto-frictional method.

For the initial density distribution of MHD simulations, we first define a piecewise function for temperature $T(r)$:
\begin{equation}
\label{mhd_temperature}
    T=\left\{\begin{aligned}
    &T_0 &&{r_{min}\leq r \leq r_0,}\\
    &k_T(r-R_0)+T_0 &&{r_0\leq r \leq r_1,}\\
    &T_1 &&{r_1\leq r \leq r_{max},}
    \end{aligned}\right.
\end{equation}
where $T_0=8.5\times 10^3$ K, $T_1=1.0\times 10^6$ K, $r_{min}=1.001R_\sun$, $r_{max}=1.801R_\sun$, $r_0=1.005R_\sun$, $r_1=1.014R_\sun$, and $k_T$ is a linear coefficient. We use this distribution to compute the initial density $\rho_{init}(r)$ based on the hydrostatic condition $\frac{dp}{dr}=-g\rho_{init}$
where $p=\rho_{init} T$. The density distribution $\rho_{init}(r)$ is then derived combined with Equations (\ref{mhd_temperature}). The normalized density unit $\rho_0=2.34\times10^{-15}~\mathrm{g}~\mathrm{cm}^{-3}$, and the density on the solar surface $r=1R_\sun$ is $\rho(R_\sun)=1.9\times10^8\rho_0$. To keep the embedded MFR in a steady state as observed during the early evolution phase, we modify the density within the computational domain during the first 28 minutes of the simulation, from 04:00 to 04:28 UT. We apply a similar manipulation as in \citet{guoDataconstrainedMagnetohydrodynamicSimulation2021, guo2023data}. It is well accepted that filament drainage can result in the non-equilibrium and thereby triggering its onset \citep{Jack2019}. Based on such a scenario, the filament eruption is triggered by adjusting the density distribution to mikic the filament drainage. The initial density $\rho(r)$ is increased to $1.0\times 10^5\rho_0$ where the density is smaller than this value, namely $\rho(r)=\max(\rho_{init},10^5\rho_0)$. From 04:29 to 04:53 UT, the initial density at 04:29 UT is reset to the stratified coronal density profile $\rho_{init}$ to simulate the drainage of the filament. This adjustment of the density is aimed to reproduce typical characteristics (rise time, eruption behavior, etc) in the evolution process of the filament with the zero-$\beta$ MHD model. Moreover, the initial velocity in the computational domain is zero everywhere. 

For the boundary condition of the density and velocity, we adopt the data-constrained case described in \citet{guo2019solar} with 2 ghost layers. As for the magnetic field, the inner ghost layer near the physical domain is fixed to the initial magnetic field data, while the outer ghost layer farther from the physical domain employs a one-sided 2nd order constant value extrapolation. Moreover, we set an artificial resistivity $\eta$ in the last 7 minutes of the high-density evolution, namely 04:21--04:28 UT. This is to better dissipate the twist of the MFR and to control magnetic reconnection, which can mitigate the kink instability and non-radial rotation of the MFR, and makes the simulation more consistent with the observed radial ejection. The artificial resistivity $\eta$ is:
\begin{gather}
    \eta=\eta_0[(J-J_c)/J_c]^2,
\end{gather}
where $J > J_c$, and we set $\eta_0=8.099\times10^{13}~\mathrm{cm}^2~\mathrm{s}^{-1}$, $J_c=1.029\times 10^{-9}~\mathrm{A\cdot cm}^{-2}$. It is turned off at later times to keep the coherent shape of the MFR.

For the numerical methods of the simulation, we employ the Strong Stability Preserving Runge-Kutta 3rd order (SSPRK3) method \citep{Gottlieb1998TotalVD} for a three-step time integration, which satisfies the Total Variation Diminishing (TVD) condition and allows the Courant–Friedrichs–Lewy (CFL) condition to reach 1. We use the Harten-Lax-van Leer (HLL) solver and the Koren limiter as the slope limiter. 

\subsection{Results and Analysis}
\label{sec:results_analysis}
\subsubsection{Simulation Results}
\label{sec:simulation results}
\begin{figure*}
    \centering
    \includegraphics[width=0.8\textwidth]{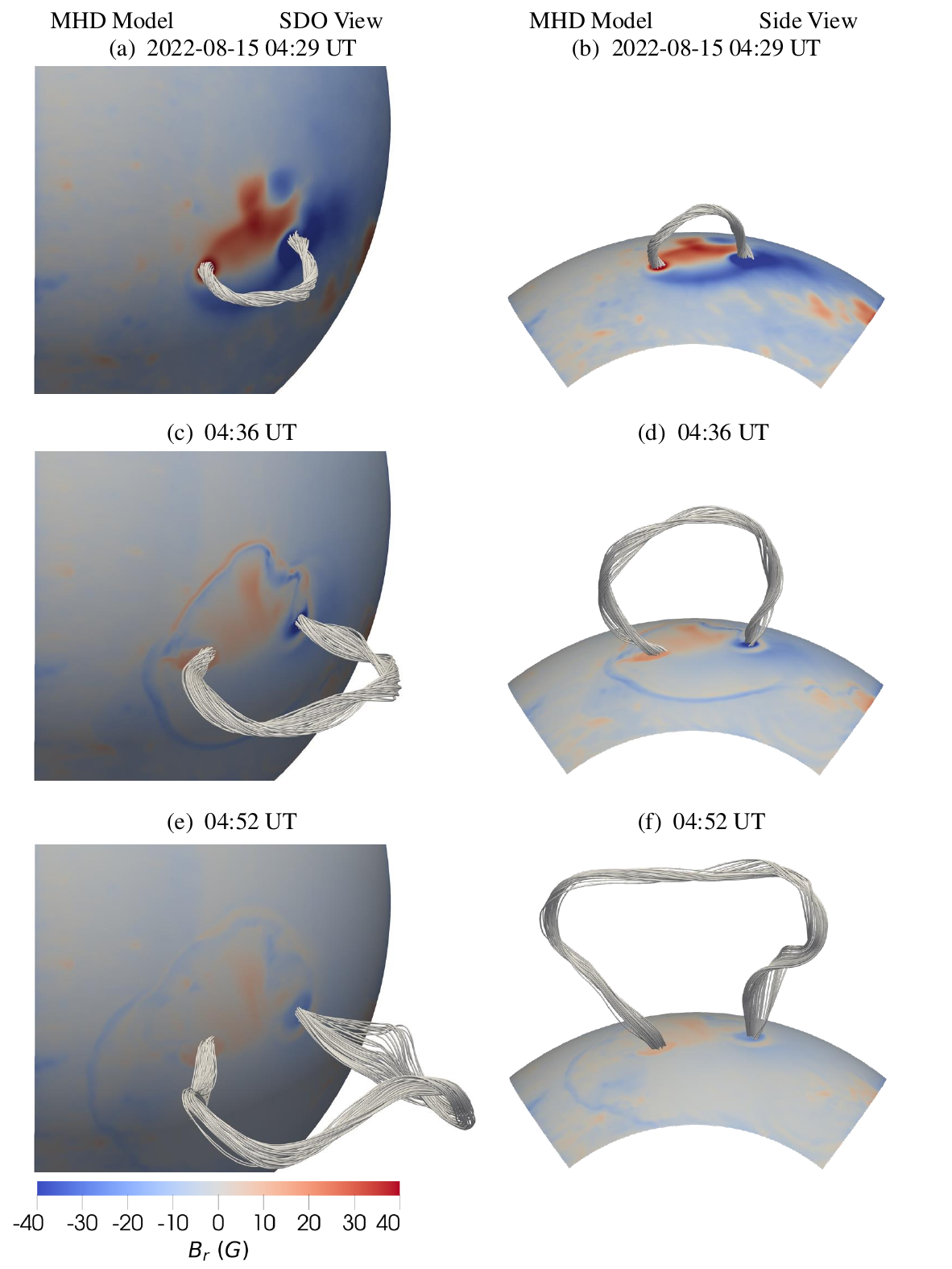}
    \caption{Evolution process of the MFR during the MHD simulation. (a), (c), and (e) SDO view at 04:29 UT, 04:36 UT, and 04:52 UT, with $B_r$ on $r=1.05R_\odot$ plane. (b), (d), and (f) Side view. An animation showing the evolution of the simulation from 04:29 to 04:52 UT is available online. }
    \label{fig:mhd_simulation1}
\end{figure*}

The zero-$\beta$ MHD simulation shows that the MFR remains stable during 04:00--04:21 UT with no significant variations in shape, height, and positions of footpoints, which is similar to observations. After we set the artificial resistivity at 04:21 UT, magnetic diffusion happens in regions with strong currents and magnetic reconnections increase numerical dissipation. The embedded MFR becomes more consistent with the filament in a decaying weak active region. At 04:29 UT, as the filament drainage happens, the axis of the MFR begins to rise, while the body of the MFR expands radially outward. Then at 04:53 UT, the MFR has widely expanded and risen up to a height that is close to the top boundary in the computation domain. We show three different times at 04:29, 04:36, and 04:52 UT both at the SDO view and a side view in Figure \ref{fig:mhd_simulation1} to show this acceleration phase. The MFR from RBSL+PFSS+NLFFF further relaxed through artificial resistivity is shown in Figures \ref{fig:mhd_simulation1}a, \ref{fig:mhd_simulation1}b. Then it starts to rise up and accelerate as in Figures \ref{fig:mhd_simulation1}c, \ref{fig:mhd_simulation1}d, \ref{fig:mhd_simulation1}e, and \ref{fig:mhd_simulation1}f. \textbf{It is noted that the bottom plane in Figure \ref{fig:mhd_simulation1} displays $B_r$ on $r=1.05R_\odot$ surface, which exhibits dissipation and evolution through the simulation process.} 

\begin{figure*}
    \centering
    \includegraphics[width=0.8\textwidth]{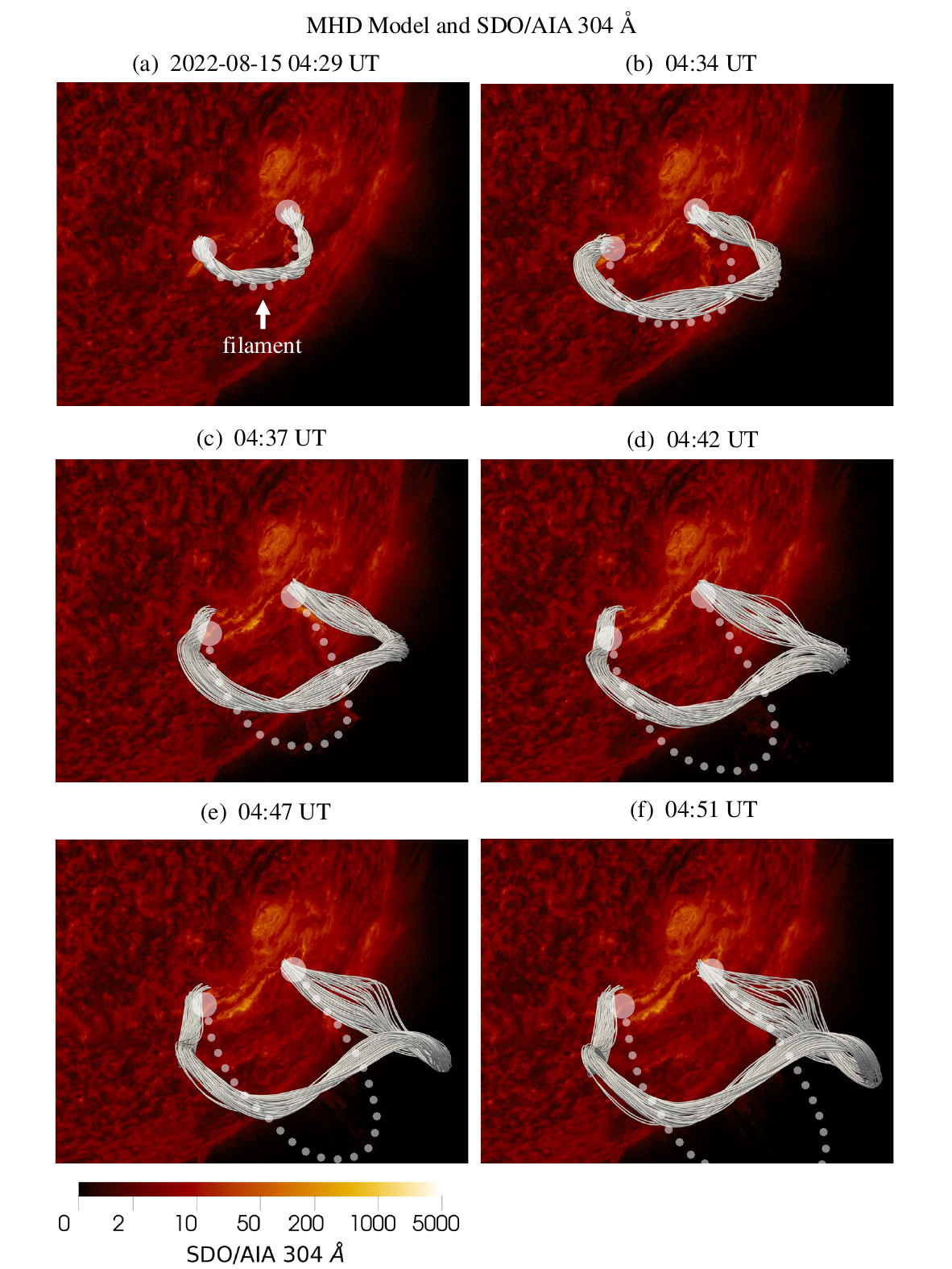}
    \caption{MHD simulation overlaid on SDO/AIA 304 \AA \ images. (a)--(f) The evolution of the MFR and corresponding 304 \AA \ observation at 04:29, 04:34, 04:37, 04:42, 04:47, and 04:51 UT. The white dotted lines represent the path of the filament in 304 \AA.}
    \label{fig:mhd_simulation2}
\end{figure*}

We also compare simulation results with SDO/AIA 304 \AA\ observations in Figure \ref{fig:mhd_simulation2}. We align the MFR field lines with 304 \AA \ images and six different evolution times are shown. The white dotted line in every panel indicates the shape of the observed filament, with big white dots displaying  positions of footpoints. At 04:29 UT shown in Figure \ref{fig:mhd_simulation2}a, the initial MFR coincides with the filament in morphology. After that at 04:34, 04:27, and 04:42 UT as shown in Figures \ref{fig:mhd_simulation2}b, \ref{fig:mhd_simulation2}c, and \ref{fig:mhd_simulation2}d, both the simulated MFR and observed filament begin to lift up and expand radially. Their eruption direction and lifting height show a good alignment. At 04:47 UT in Figure \ref{fig:mhd_simulation2}e, the MFR begins to rotate. One possible rotation mechanism is that the high twist number of the MFR in the RBSL model might trigger the kink instability, which forces the rotation. The filament has risen close to the upper boundary of the computation box with clear brightened flare ribbons. At 04:51 UT in Figure \ref{fig:mhd_simulation2}f, the simulated MFR gets close to the upper boundary while the observed filament has risen outside the field of view. 

From the comparison between simulation results and observations, we can conclude two key points. First, the embedded MFR after magneto-frictional relaxation is highly consistent with the observed filament, indicating that the parameters for the RBSL model, including the path $\mathcal{C}$, minor radius $a$, and flux $F$, such that they recover well the observational evolution. Secondly, the MFR matches the filament in the early phase of the eruption. Therefore, it is possible to reproduce the triggering and eruption of the filament with zero-$\beta$ MHD simulations in spherical coordinates.

\subsubsection{Kinematics}
\label{sec:kinematics}
\begin{figure*}
    \centering
    \includegraphics[width=0.8\textwidth]{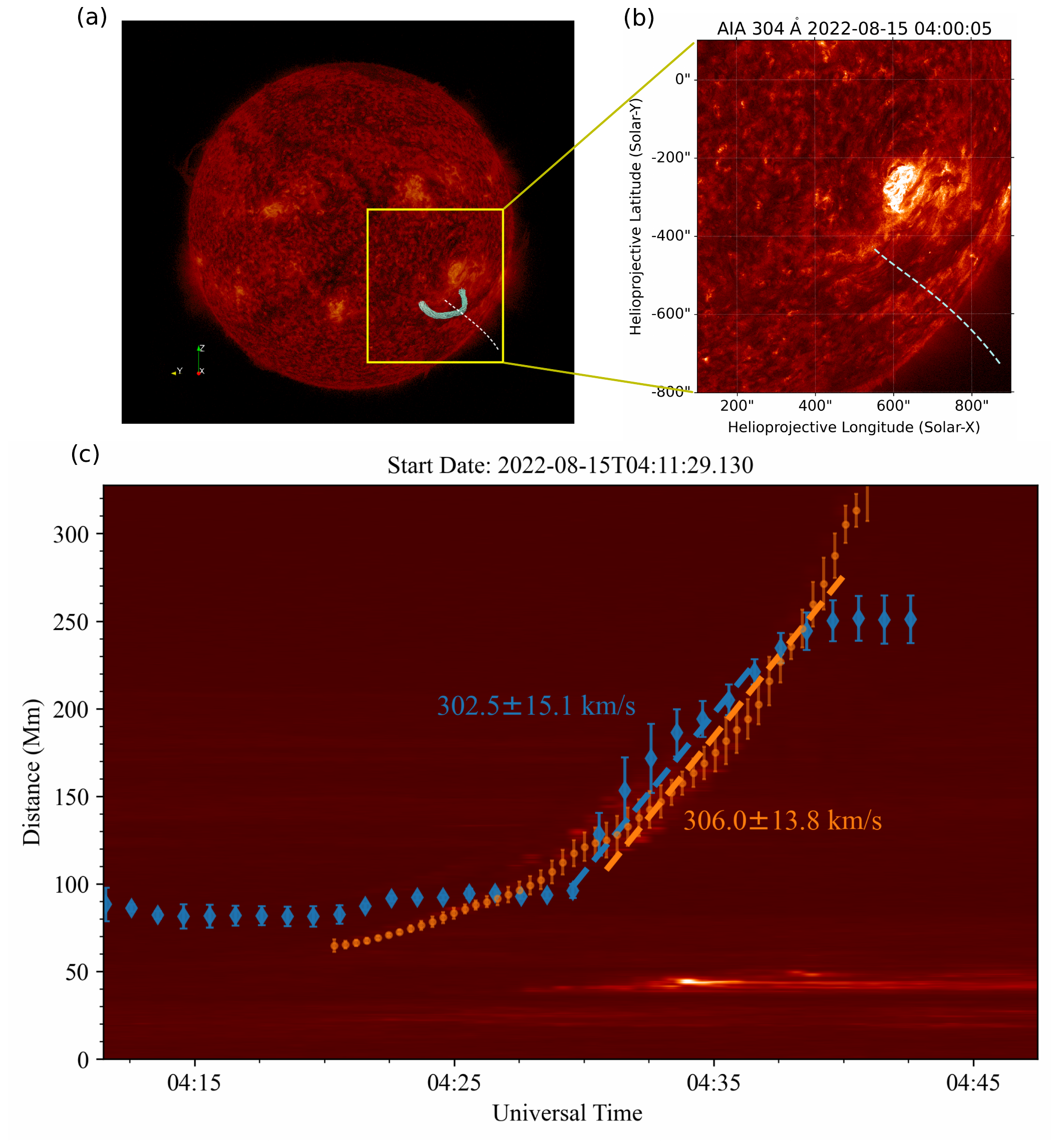}
    \caption{Kinematics of the filament in observation and the MFR in MHD simulation. (a) The MFR from MHD simulation overlaid on the SDO/AIA 304 \AA \ image at 04:00 UT. The yellow square shows the domain that we choose to analyze the time-distance profile of the filament. The white dashed line shows the slice path that we choose to measure the velocity of both the filament from observations and the erupting MFR from simulations. (b) The enlarged 304 \AA \ image shown in the yellow square in panel (a). (c) Time-distance diagram of the 304 \AA \ images and height-time profiles derived from both observations and simulations. Orange dots and vertical line segments represent the positions and errors measured from the observation of SDO/AIA 304 \AA \ waveband. Blue dots and vertical line segments represent the positions and errors measured from the MHD simulation. The orange dashed line is a linear fitting to the eruption velocity of the observed filament, deriving a velocity of $306.0~\mathrm{km}~\mathrm{s}^{-1}$. The blue dashed line represents similar results from simulations, deriving a velocity of $302.5~\mathrm{km}~\mathrm{s}^{-1}$.}
    \label{fig:mhd_time_distance}
\end{figure*}

We analyze the kinematics in our simulations to compare the results quantitatively with observations. The time-distance profiles are shown in Figure \ref{fig:mhd_time_distance}. We select a slice path both for SDO/AIA 304 \AA \ images and simulation snapshots, which is displayed as the white dashed line in Figures \ref{fig:mhd_time_distance}a and \ref{fig:mhd_time_distance}b. This path is tracked along the highest position of the filament material frame by frame on running difference images of SDO/AIA 304 \AA \ images. Then, we measure axis positions of the ejected filament and the MFR during the eruption, which is repeated five times to decrease the measurement errors. Time-distance profiles are displayed in Figure \ref{fig:mhd_time_distance}c. We find that they both have a stable phase with no large changes in position and shape, and a rapid-rise phase with a fast lifting speed. In the rapid-rise phase, the observed ejection velocity of the filament is $306.0\pm13.8 ~\mathrm{km}~\mathrm{s}^{-1}$, while that of the MFR is $302.5\pm15.1~ \mathrm{km}~\mathrm{s}^{-1}$. They are consistent within the error range, indicating that our simulation reproduces the kinematics of the filament eruption. It is noted that in the late stage of simulations, the MFR cannot continue to rise as observed due to the MFR lifting to the upper boundary and the limitations of the RBSL method and zero-$\beta$ MHD model. Actually, the ejection velocity of the simulated MFR is slower than the observed one after $\sim$ 04:38 UT. 

\begin{figure*}
    \centering
    \includegraphics[width=0.8\textwidth]{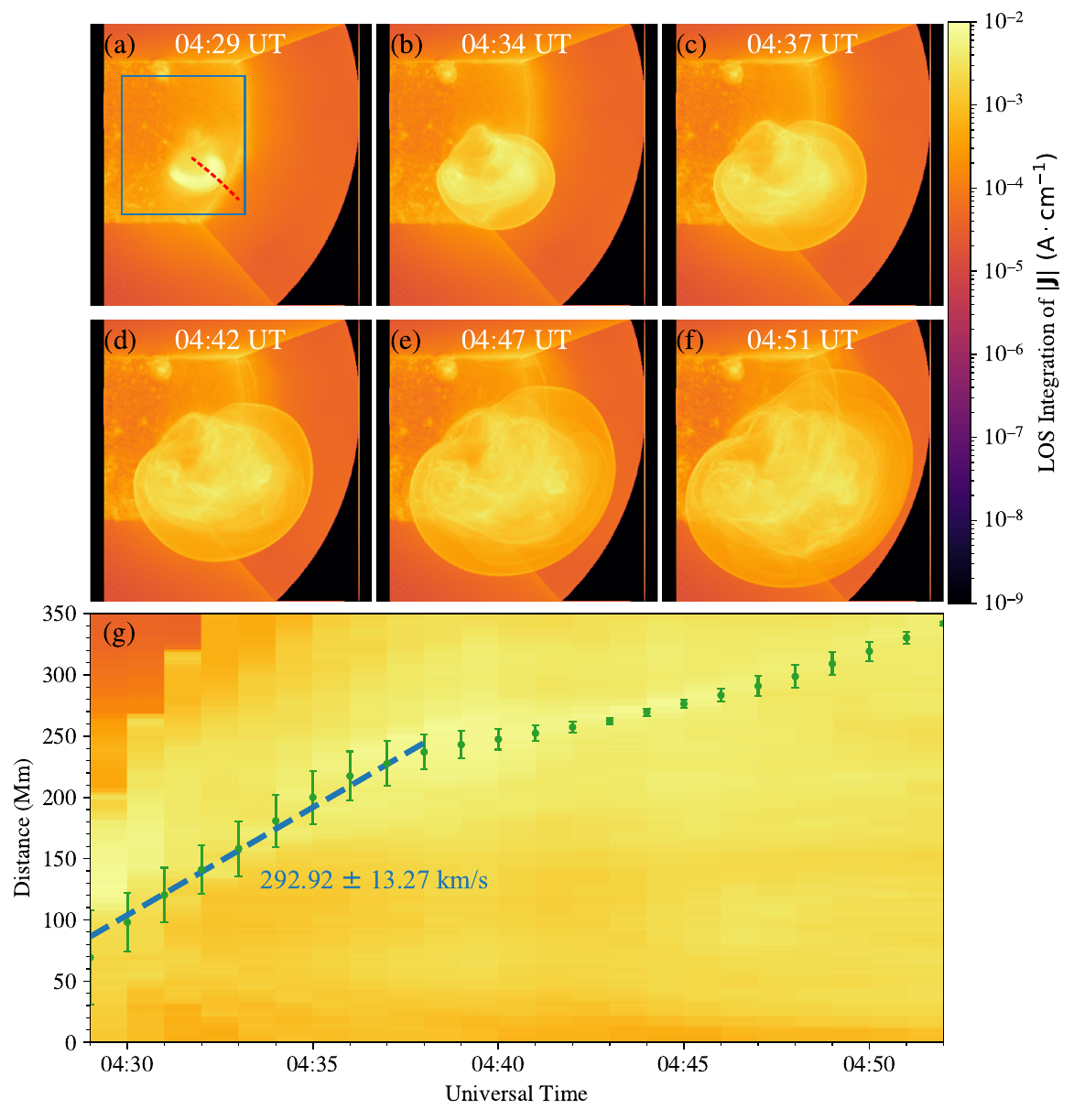}
    \caption{LOS current intensity synthesized from simulation results in SDO view and the time-distance profile calculated from the synthetic intensity. (a)--(f) LOS current intensity at 04:29, 04:34, 04:37, 04:42, 04:47, and 04:51 UT. (a) The blue square indicates the area selected for slicing, and the red dashed line represents the slice path. (g) Time-distance profile of the LOS current density. Blue dots and vertical line segments represent the positions and errors. The blue line is a linear fitting to the eruption velocity of the most intense current density, deriving a velocity of $292.9~\mathrm{km}~\mathrm{s}^{-1}$. }
    \label{fig:mhd_current_td}
\end{figure*}

Moreover, to perform a more accurate quantitative comparison, we read the simulation data in Python with the 
astrophysical data visualization open-source package Python.yt \citep{turk2011pythonyt}, and resample it from spherical coordinates to Cartesian coordinates using the Radiation Synthesis Tools\footnote{https://github.com/gychen-NJU/radsyn\_tools} (RST), which is a visualization code able to analyze data simulated by MPI-AMRVAC with high computational efficiency, also applicable to other relevant data. Due to limitations of zero-$\beta$ MHD model, which has no information on temperatures, we perform pseudo-radiation synthesis with the LOS integral of current density with the RST code. This approach is similar to other published works. For example, \citet{warnecke2019data} proposed Ohmic dissipation as a major heating source during coronal structure formation, which indicates that we can use current density for synthesized pseudo-radiation. \citet{jarolim2023probing} also employed a similar method to compare observations with force-free field models containing only magnetic field information. Here, we integrate the current density along the line of sight in SDO view and synthesized pseudo-radiation images are shown in Figure \ref{fig:mhd_current_td}. We select the same region and slice path as Figure \ref{fig:mhd_time_distance} and measure the eruption velocity of the most intense current area around the MFR, which can denote the lifting of the observed filament spine. Figures \ref{fig:mhd_current_td}a--\ref{fig:mhd_current_td}f display six different times during the simulation at 04:29, 04:34, 04:37, 04:42, 04:47, and 04:51 UT. The MFR has expanded and lifted up, with a clearly visible expanded current contour and an uplift axis exhibiting strong current. Figure \ref{fig:mhd_current_td}g is the time-distance profile of the pseudo-radiation. It can be seen that the rise velocity of the axis during the eruption phase is $292.92\pm13.27 ~\mathrm{km}~\mathrm{s}^{-1}$, which is consistent with the eruption velocity of the observed filament and that obtained from slicing the magnetic field lines of the MFR, within the error range. Furthermore, in order to quantitatively analyze the density adjustment in the simulation, specifically the triggering effect of filament drainage on the eruption of magnetic flux ropes, we calculate the acceleration exerted by the radial Lorentz force, namely $a_L=|(\mathbf{J}\times\mathbf{B})_r|/\rho$, on the MFR both before and after the occurrence of filament drainage, which is relatively $a_L=0.6141~\mathrm{m/s^2}$ at 04:28 UT and $a_L=~861.4~\mathrm{m/s^2}$ at 04:29 UT, calculating by averaging 50 magnetic field lines of the MFR. This suggests that under the influence of radial Lorentz forces of approximately equal magnitude, the drainage of filament material leads to a larger radial acceleration of the MFR, leading it to lose equilibrium and erupt.

\subsubsection{Magnetic Topology}
\label{sec:topology}
\begin{figure*}
    \centering
    \includegraphics[width=0.8\textwidth]{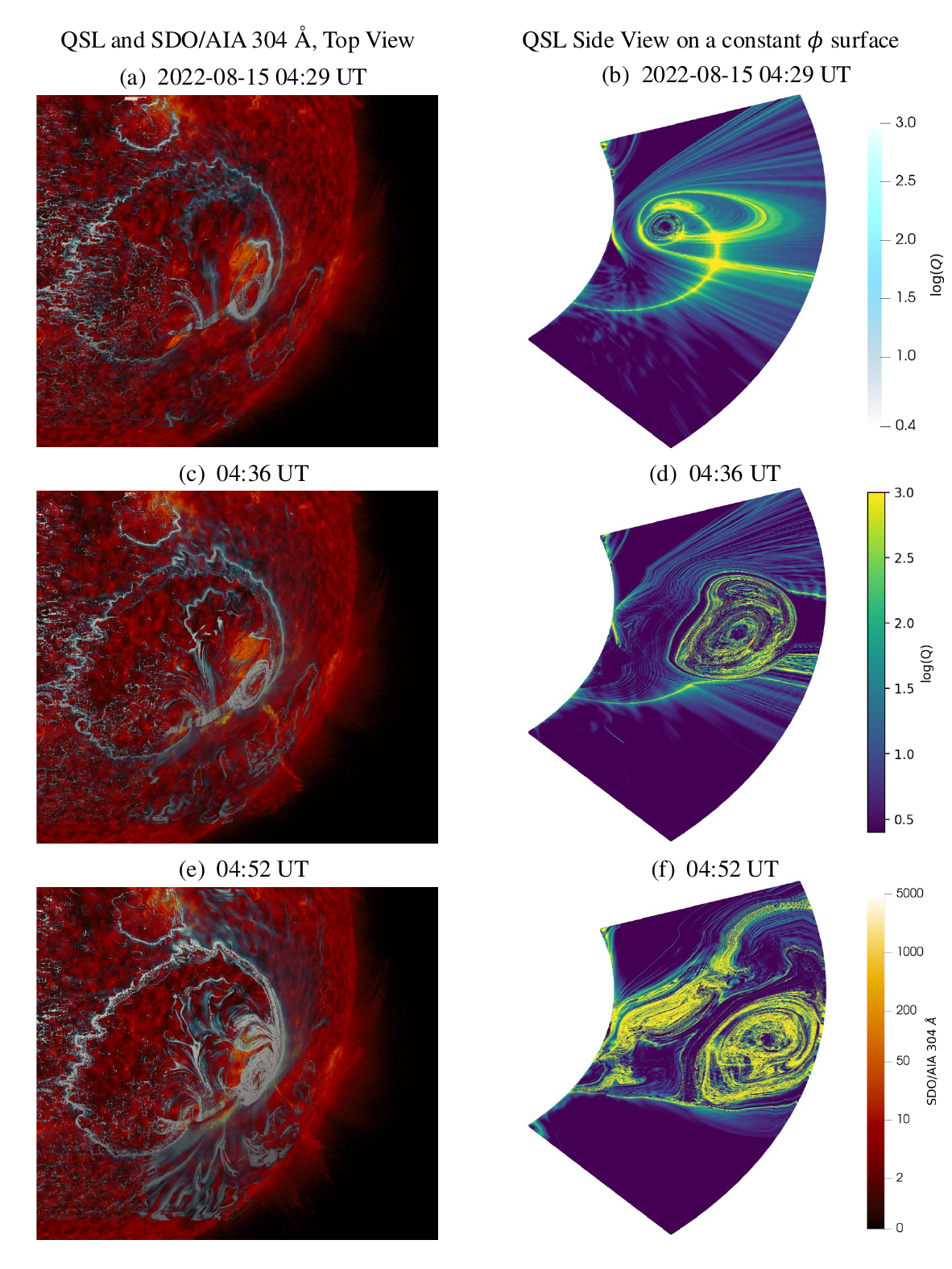}
    \caption{Distribution of QSLs calculated from simulation results. (a), (c), and (e) Distribution of $\log(Q)$ at the $r=1.001R_s$ bottom boundary at 04:29 UT, 04:36 UT, and 04:52 UT, overlaid on observations from SDO/AIA 304 Å. (b), (d), and (f) Distribution of $\log(Q)$ on the $\phi=222.52\degree$ plane.}
    \label{fig:mhd_qsl}
\end{figure*}

Qusai-Separatrix Layers (QSLs) represent regions where the magnetic field line linkages change rapidly. QSLs are defined by the squashing factor $Q \gg 2$ \citep{demoulin1996quasi, titov2002theory}, and possess complex magnetic field characteristics, where small disturbances can lead to considerable changes in magnetic topology. These regions are also where magnetic reconnection is likely to happen. There are multiple methods to compute QSLs \citep{titovGeneralizedSquashingFactors2007, pariatEstimationSquashingDegree2012, tassevQSLSquasherFast2017a, scott2017magnetic}. We calculate QSLs for both bottom surface at $r=1.001R_\sun$ and the vertical plane intersecting the axis of the MFR at $\phi=222.52\degree$ using the K-QSL code in spherical coordinates \citep{https://doi.org/10.5281/zenodo.13831036}, illustrated in Figure \ref{fig:mhd_qsl}, which displays $\log(Q)$ at 04:29, 04:36, and 04:52 UT.

With QSLs on the bottom surface overlaid on SDO/AIA 304 \AA\ images, we find that MFR footpoint positions have relative high $Q$ values. Moreover, there are QSLs coinciding with the flare ribbon at 304 \AA \ images. This indicates multiple magnetic reconnections occur inside the eruption MFR and at the hyperbolic flux tube below the MFR. During the period of 04:42--04:52 UT, QSLs near the flare ribbon display a stripe-like separation, consistent with the observed separation of the observed flare ribbons. With QSLs on the vertical plane as shown in Figures \ref{fig:mhd_qsl}b, \ref{fig:mhd_qsl}d, and \ref{fig:mhd_qsl}f, we find that strong $\log(Q)$ values appear in the cross-section of the MFR, which gradually lifts and expands.

\begin{figure*}
    \centering
    \includegraphics[width=0.5\textwidth]{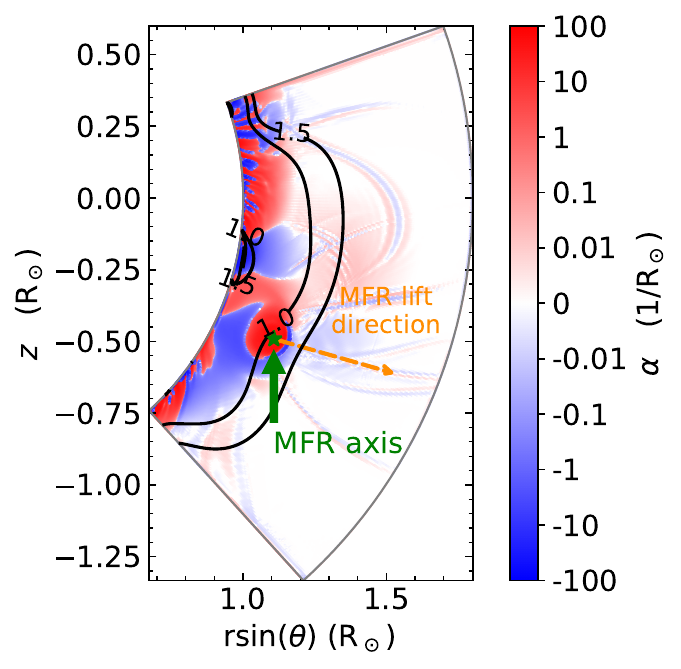}
    \caption{Twist rate $\alpha$ in the cross-section of the MFR axis midpoint, over-plotted with contours of the decay index of the horizontal potential field. The cross-section center is marked by a green star, with a green arrow indicating its position. The orange dashed line denotes the MFR lift direction.}
    \label{fig:mhd_twist_decay}
\end{figure*}

Moreover, to further investigate the triggering mechanism of the MFR, we calculate the distribution of the twist rate $\alpha=\mathbf{J}\cdot\mathbf{B}/B^2$ and the decay index of the external horizontal potential field $\mathrm{d}(\mathrm{ln}P_h)/\mathrm{d}(\mathrm{ln}h)$ in the cross-section plane cross the middle position of the MFR axis at the start time of the eruption, 04:29 UT, shown in Figure \ref{fig:mhd_twist_decay}. This analytical approach is based on \citet{fanImprovedMagnetohydrodynamicSimulation2022,2024ApJFan}. It can be seen that the MFR axis exhibits a large twist rate and that the majority of the MFR axis are located below the contour of a decay index of 1.5. We also calculate the total twist by integrating the twist rate along the selected field lines and it is found that the average total twist of the field lines between the footpoints reaches approximately 2 winds, which exceeds the critical total twist of 1.25 winds, indicating the onset of kink instability. However, as indicated by the eruption direction of the magnetic flux rope axis marked in Figure \ref{fig:mhd_twist_decay}, during the subsequent eruption process, the MFR continues to rise above the contour of a decay index of 1.5, suggesting that torus instability is likely to also participate in and influence its eruption. Therefore, it indicates that the initial lifting of the MFR is primarily driven by kink instability, with the torus instability involving in its later stages.

\section{Summary and Discussion} \label{sec:summary}
Combining the RBSL and PFSS models, we reconstruct the coronal magnetic field of AR 13079 in spherical coordinates, which experienced a C3.5 flare triggered by a filament eruption on 2022 August 15. After using a magneto-frictional method to relax the three-dimensional magnetic field, we conduct a zero-$\beta$ MHD simulation in spherical coordinates and successfully reproduced the filament eruption, consistent with SDO/AIA observations in morphology, eruption timing, lifting velocity, and eruption direction.

To analyze simulation results and compare them with observations, we first directly compare the morphology of the MFR with the observed filament and find a good agreement in positions of footpoints and shape \textbf{as shown in Figures \ref{fig:mhd_simulation1} and \ref{fig:mhd_simulation2}}. \textbf{With the prescribed bottom boundary conditions, the radial magnetic field in the inner ghost layer is fixed to the observed magnetic field, thereby satisfying the line-tied condition on the bottom boundary. Therefore, it is noteworthy that a distinct evolution of $B_r$ at the surface $r=1.05R_\odot$ in Figure \ref{fig:mhd_simulation1} demonstrates that pronounced current layers appear during the eruptive phase. In the future, we can attempt to detect such current layers, which would imply that the line-tying effects remain valid in photosphere.} Secondly, we calculate time-distance profiles of SDO/AIA 304 \AA \ observations, integrate magnetic field lines and synthesize pseudo-radiation images. The eruption timings and velocities in rapid-rise phase from these plots are highly consistent and our simulations reproduce the eruption speed of $302.5\pm15.1 ~\mathrm{km}~\mathrm{s}^{-1}$ in Figure \ref{fig:mhd_time_distance}c. Thirdly, we calculate the QSLs on the bottom plane and the $\phi=222.52\degree$ plane. \textbf{By quantitatively analyzing the eruption process, we find that the high-Q lines at the bottom surface denote the footpoint positions of those field lines where the magnetic reconnection can likely occur from Figures \ref{fig:mhd_qsl}(a), \ref{fig:mhd_qsl}(c), and \ref{fig:mhd_qsl}(e)}. On the vertical plane, the expansion and lifting of QSLs around the MFR indicates the evolution of the filament \textbf{from Figures \ref{fig:mhd_qsl}(b), \ref{fig:mhd_qsl}(d), and \ref{fig:mhd_qsl}(f)}. Therefore, our simulations faithfully reproduce the triggering and eruption of this eruptive filament. 

We find a good agreement between our simulation and observations in kinematics and magnetic topology, indicating that the RBSL+PFSS model after magneto-frictional extrapolation is an applicable method to reconstruct the three-dimensional coronal magnetic field of filaments in weak and decaying active regions in spherical coordinate system. With the RBSL method, we can qualitatively identify the most important parameters of the embedded MFR with the eruption mechanism. First, the toroidal flux of the MFR is a key factor for the eruption. If the flux $F$ is too small, i.e., $F_0$, $2F_0$, $4F_0$, $6F_0$, etc., the MFR will sink below the solar surface instead of successfully erupting. If the flux is too large, the MFR will rise with large rotation in MHD simulations because of the large twist number. We ultimately determine an appropriate flux that ensures the initial configuration of the MFR is consistent with observations while maintaining an accurate eruption velocity and other dynamic characteristics in subsequent simulations. Secondly, the MFR tends to erupt successfully without significant rotation with a minor radius $a$ ranging in 4--5 times of the observed filament radius and a suitable flux, which confirms conclusions in \citet{guoProminenceFineStructures2022a}. Moreover, although the zero-$\beta$ model does not have temperature and pressure variations, it can be used to simulate the filament eruption in a large region in spherical coordinates. 

However, our methods have limitations. First, we employ zero-$\beta$ MHD simulations, which save computational resources but neglect thermal pressure, gravity, and the energy equation. While these effects are assumed to be not dominant for this event, zero-$\beta$ MHD models neglect certain physical mechanism, such as the fact that the filament itself is cool and dense, embedded in the bottom of a hot coronal MFR. There are studies implementing full MHD simulations in spherical coordinates \citep{fanMHDSimulationsEruption2017,fanImprovedMagnetohydrodynamicSimulation2022}. What is noted that gravity and the weight of the filament material is important in realistic physical scenarios. Due to the high density of the filament relative to surroundings, its material may also fall back during the eruption process and trigger solar eruptions \citep{jenkins2018understanding, jenkins2019modeling}, which also delay the occurrence of some eruptions. Secondly, our simulation grids are uniform, and we should perform future follow-up studies with adaptive mesh refinement to capture important dynamical details near the erupting MFR. Furthermore, due to the weak magnetic field, the high twist of the RBSL model has a more pronounced impact compared to typical active regions. The embedding of the MFR has a greater effect on the background field, and the impact of adjusting the flux is also more significant. Additionally, we set the top boundary in $r$ direction at $1.801R_\odot$, and employ a second order zero-gradient extrapolation for the boundary condition. Therefore, during the later stage of the MFR eruption, the simulated MFR experiences an unreasonable deceleration phase in kinematics at the top boundary seen in Figures \ref{fig:mhd_time_distance}c and \ref{fig:mhd_current_td}g, which is inconsistent with observations. This can be remedied in future work by implementing certain absorbing boundary treatment that avoids wave reflections, or by doing the simulation in even larger radial domains.

\citet{guo2023data} also conducted a zero-$\beta$ MHD simulation in spherical coordinates to reproduce a filament eruption. This work presents further innovations and improvements in the technical approach. First, active region 13079 covers a large region, which makes the prepocessing of the vector magnetic field data more complicated. We convert the data from the local Cartesian coordinates to the Stonyhurst Heliographic coordinates and thus construct a more reliable synoptic frame. Secondly, a large active region results in a lower simulation resolution, such that our fixed grid simulation does experience significant numerical dissipation. To remedy this, we experiment with the choices offered in MPI-AMRVAC to vary numerical schemes and parameters to make the simulation more reliable. Thirdly, this active region is decaying with relatively weak magnetic field intensity, which makes it more difficult to reproduce the filament eruption, particularly in ensuring that the MFR remains consistent with observations in its initial static configuration while also matching its consequent dynamic evolution. Consequently, both the magnetic field reconstruction and MHD simulations require multiple parameter testings to achieve a more realistic and accurate eruption. Fourthly, compared to \citet{guo2023data}, the setting of artificial resistivity and density distribution is different in this study. In order to perform a more realistic MHD relaxation, we test a set of different parameters that control our anomalous resistivity, namely $\eta_0$, $J_c$, and relaxation time and finally determine parameters which match this weak decaying active region. Moreover, the MHD relaxation with high resistivity is implemented after the high-density evolution, which makes the timing of magnetic reconnections more physically accurate. Finally, in this work, a more comprehensive approach is adopted for the analysis of simulation results. We combine the time-distance profiles, QSLs distributions, and the pseudo-radiation synthesis with the simulated current density to conduct a quantitative analysis in terms of morphology and kinematics. In combination, this approach allows reconstructing, simulating, and analyzing filament eruptions in decaying active regions. 

Recent studies have successfully integrated MFR eruptions with global coronal magnetic field, solar wind models, and space weather effects \citep{verbekeICARUSNewInner2022,perri2022coconut,lionelloGlobalMHDSimulations2023}. In the future, we can further extend to even more global domains, and realize fully space-weather relevant global coronal magnetic field simulation.

\acknowledgments

The observational data are provided courtesy of NASA/SDO and the AIA and HMI science teams. Y.H.L., Y.G., J.H.G., and M.D.D. are supported by the National Key R\&D Program of China (2022YFF0503004, 2021YFA1600504, and 2020YFC2201200) and NSFC (12333009, 123B1025). R.K. received funding from the European Research Council (ERC) under the European Union’s Horizon 2020 research and innovation program (grant agreement No. 833251 PROMINENT ERC-ADG 2018), and from Internal funds KU Leuven, project C14/19/089 TRACESpace and FWO project G0B4521N. The numerical simulations are performed in the High Performance Computing Center (HPCC) at Nanjing University.


\bibliographystyle{apj}
\bibliography{biblo_20220815}

\begin{thebibliography}{74}
\expandafter\ifx\csname natexlab\endcsname\relax\def\natexlab#1{#1}\fi

\bibitem[{An(1985)}]{an1985formation}
An, C.-H. 1985, Astrophysical Journal, Part 1 (ISSN 0004-637X), vol. 298, Nov. 1, 1985, p. 409-413. Research supported by the National Research Council and NADA., 298, 409

\bibitem[{{Antiochos} \& {Klimchuk}(1991)}]{1991antiochosfilament}
{Antiochos}, S.~K. \& {Klimchuk}, J.~A. 1991, \apj, 378, 372

\bibitem[{{Chen} {et~al.}(2022){Chen}, {Rempel}, \& {Fan}}]{2022ApJChenfengRMHDflux}
{Chen}, F., {Rempel}, M., \& {Fan}, Y. 2022, \apj, 937, 91

\bibitem[{Chen {et~al.}(2014)Chen, Harra, \& Fang}]{chen2014imaging}
Chen, P., Harra, L., \& Fang, C. 2014, The Astrophysical Journal, 784, 50

\bibitem[{Chen(2011)}]{chenCoronalMassEjections2011}
Chen, P.~F. 2011, Living Rev. Solar Phys., 8

\bibitem[{{Chen} {et~al.}(2020){Chen}, {Xu}, \& {Ding}}]{Chen2020}
{Chen}, P.-F., {Xu}, A.-A., \& {Ding}, M.-D. 2020, Research in Astronomy and Astrophysics, 20, 166

\bibitem[{Cheng {et~al.}(2017)Cheng, Guo, \& Ding}]{cheng2017origin}
Cheng, X., Guo, Y., \& Ding, M. 2017, Science China Earth Sciences, 60, 1383

\bibitem[{Dalmasse {et~al.}(2019)Dalmasse, Savcheva, Gibson, Fan, Nychka, Flyer, Mathews, \& DeLuca}]{dalmasse2019data}
Dalmasse, K., Savcheva, A., Gibson, S., Fan, Y., Nychka, D., Flyer, N., Mathews, N., \& DeLuca, E. 2019, The Astrophysical Journal, 877, 111

\bibitem[{Demoulin {et~al.}(1996)Demoulin, Henoux, Priest, \& Mandrini}]{demoulin1996quasi}
Demoulin, P., Henoux, J.-C., Priest, E., \& Mandrini, C. 1996, Astronomy and Astrophysics, v. 308, p. 643-655, 308, 643

\bibitem[{Fan(2001)}]{fan2001emergence}
Fan, Y. 2001, The Astrophysical Journal, 554, L111

\bibitem[{Fan(2017)}]{fanMHDSimulationsEruption2017}
---. 2017, ApJ, 844, 26

\bibitem[{Fan(2022)}]{fanImprovedMagnetohydrodynamicSimulation2022}
---. 2022, ApJ, 941, 61

\bibitem[{{Fan} {et~al.}(2024){Fan}, {Kazachenko}, {Afanasyev}, \& {Fisher}}]{2024ApJFan}
{Fan}, Y., {Kazachenko}, M.~D., {Afanasyev}, A.~N., \& {Fisher}, G.~H. 2024, \apj, 975, 206

\bibitem[{Gilbert {et~al.}(2007)Gilbert, Alexander, \& Liu}]{gilbert2007filament}
Gilbert, H.~R., Alexander, D., \& Liu, R. 2007, solar physics, 245, 287

\bibitem[{Gottlieb \& Shu(1998)}]{Gottlieb1998TotalVD}
Gottlieb, S. \& Shu, C.-W. 1998, Math. Comput., 67, 73

\bibitem[{Guo {et~al.}(2023{\natexlab{a}})Guo, Ni, Zhong, Guo, Xia, Li, Poedts, Schmieder, \& Chen}]{guo2023thermodynamic}
Guo, J., Ni, Y., Zhong, Z., Guo, Y., Xia, C., Li, H., Poedts, S., Schmieder, B., \& Chen, P. 2023{\natexlab{a}}, The Astrophysical Journal Supplement Series, 266, 3

\bibitem[{Guo {et~al.}(2022)Guo, Ni, Zhou, Guo, Schmieder, \& Chen}]{guoProminenceFineStructures2022a}
Guo, J., Ni, Y., Zhou, Y.-H., Guo, Y., Schmieder, B., \& Chen, P. 2022, Astronomy \& Astrophysics, 667

\bibitem[{Guo {et~al.}(2021{\natexlab{a}})Guo, Zhou, Guo, Ni, Karpen, \& Chen}]{guo2021formation}
Guo, J., Zhou, Y., Guo, Y., Ni, Y., Karpen, J., \& Chen, P. 2021{\natexlab{a}}, The Astrophysical Journal, 920, 131

\bibitem[{Guo {et~al.}(2017)Guo, Cheng, \& Ding}]{guoOriginStructuresSolar2017}
Guo, Y., Cheng, X., \& Ding, M. 2017, Sci. China Earth Sci., 60, 1408

\bibitem[{Guo {et~al.}(2023{\natexlab{b}})Guo, Guo, Ni, Ding, Chen, Xia, Keppens, \& Yang}]{guo2023data}
Guo, Y., Guo, J., Ni, Y., Ding, M., Chen, P., Xia, C., Keppens, R., \& Yang, K.~E. 2023{\natexlab{b}}, The Astrophysical Journal, 958, 25

\bibitem[{Guo {et~al.}(2016{\natexlab{a}})Guo, Xia, \& Keppens}]{guo2016magneto2}
Guo, Y., Xia, C., \& Keppens, R. 2016{\natexlab{a}}, The Astrophysical Journal, 828, 83

\bibitem[{Guo {et~al.}(2019)Guo, Xia, Keppens, Ding, \& Chen}]{guo2019solar}
Guo, Y., Xia, C., Keppens, R., Ding, M., \& Chen, P. 2019, The Astrophysical Journal Letters, 870, L21

\bibitem[{Guo {et~al.}(2016{\natexlab{b}})Guo, Xia, Keppens, \& Valori}]{guo2016magneto1}
Guo, Y., Xia, C., Keppens, R., \& Valori, G. 2016{\natexlab{b}}, The Astrophysical Journal, 828, 82

\bibitem[{{Guo} {et~al.}(2019){Guo}, {Xu}, {Ding}, {Chen}, {Xia}, \& {Keppens}}]{guo2019mfr}
{Guo}, Y., {Xu}, Y., {Ding}, M.~D., {Chen}, P.~F., {Xia}, C., \& {Keppens}, R. 2019, \apjl, 884, L1

\bibitem[{Guo {et~al.}(2021{\natexlab{b}})Guo, Zhong, Ding, Chen, Xia, \& Keppens}]{guoDataconstrainedMagnetohydrodynamicSimulation2021}
Guo, Y., Zhong, Z., Ding, M.~D., Chen, P.~F., Xia, C., \& Keppens, R. 2021{\natexlab{b}}, ApJ, 919, 39

\bibitem[{Jarolim {et~al.}(2023)Jarolim, Thalmann, Veronig, \& Podladchikova}]{jarolim2023probing}
Jarolim, R., Thalmann, J., Veronig, A., \& Podladchikova, T. 2023, Nature Astronomy, 7, 1171

\bibitem[{{Jenkins} {et~al.}(2019){Jenkins}, {Hopwood}, {D{\'e}moulin}, {Valori}, {Aulanier}, {Long}, \& {van Driel-Gesztelyi}}]{Jack2019}
{Jenkins}, J.~M., {Hopwood}, M., {D{\'e}moulin}, P., {Valori}, G., {Aulanier}, G., {Long}, D.~M., \& {van Driel-Gesztelyi}, L. 2019, \apj, 873, 49

\bibitem[{Jenkins {et~al.}(2019)Jenkins, Hopwood, D{\'e}moulin, Valori, Aulanier, Long, \& van Driel-Gesztelyi}]{jenkins2019modeling}
Jenkins, J.~M., Hopwood, M., D{\'e}moulin, P., Valori, G., Aulanier, G., Long, D.~M., \& van Driel-Gesztelyi, L. 2019, The Astrophysical Journal, 873, 49

\bibitem[{Jenkins {et~al.}(2018)Jenkins, Long, van Driel-Gesztelyi, \& Carlyle}]{jenkins2018understanding}
Jenkins, J.~M., Long, D.~M., van Driel-Gesztelyi, L., \& Carlyle, J. 2018, Solar Physics, 293, 7

\bibitem[{Kang {et~al.}(2023)Kang, Guo, Roussev, Keppens, \& Lin}]{kang2023modelling}
Kang, K., Guo, Y., Roussev, I.~I., Keppens, R., \& Lin, J. 2023, Monthly Notices of the Royal Astronomical Society, 518, 388

\bibitem[{Keppens {et~al.}(2023)Keppens, Braileanu, Zhou, Ruan, Xia, Guo, Claes, \& Bacchini}]{keppens2023mpi}
Keppens, R., Braileanu, B.~P., Zhou, Y., Ruan, W., Xia, C., Guo, Y., Claes, N., \& Bacchini, F. 2023, Astronomy \& Astrophysics, 673, A66

\bibitem[{{Keppens} {et~al.}(2003){Keppens}, {Nool}, {T{\'o}th}, \& {Goedbloed}}]{2003ronyamrvac}
{Keppens}, R., {Nool}, M., {T{\'o}th}, G., \& {Goedbloed}, J.~P. 2003, Computer Physics Communications, 153, 317

\bibitem[{{Keppens} {et~al.}(2023){Keppens}, {Popescu Braileanu}, {Zhou}, {Ruan}, {Xia}, {Guo}, {Claes}, \& {Bacchini}}]{2023ronyamrvac3.0}
{Keppens}, R., {Popescu Braileanu}, B., {Zhou}, Y., {Ruan}, W., {Xia}, C., {Guo}, Y., {Claes}, N., \& {Bacchini}, F. 2023, \aap, 673, A66

\bibitem[{Keppens {et~al.}(2021)Keppens, Teunissen, Xia, \& Porth}]{keppens2021mpi}
Keppens, R., Teunissen, J., Xia, C., \& Porth, O. 2021, Computers \& Mathematics with Applications, 81, 316

\bibitem[{Lionello {et~al.}(2023)Lionello, Downs, Mason, Linker, Caplan, Riley, Titov, \& DeRosa}]{lionelloGlobalMHDSimulations2023}
Lionello, R., Downs, C., Mason, E.~I., Linker, J.~A., Caplan, R.~M., Riley, P., Titov, V.~S., \& DeRosa, M.~L. 2023, ApJ, 959, 77

\bibitem[{Low \& Lou(1990)}]{low1990modeling}
Low, B. \& Lou, Y. 1990, The Astrophysical Journal, 352, 343

\bibitem[{Malanushenko {et~al.}(2014)Malanushenko, Schrijver, DeRosa, \& Wheatland}]{malanushenko2014using}
Malanushenko, A., Schrijver, C., DeRosa, M., \& Wheatland, M. 2014, The Astrophysical Journal, 783, 102

\bibitem[{McCauley {et~al.}(2015)McCauley, Su, Schanche, Evans, Su, McKillop, \& Reeves}]{mccauleyProminenceFilamentEruptions2015}
McCauley, P.~I., Su, Y., Schanche, N., Evans, K.~E., Su, C., McKillop, S., \& Reeves, K.~K. 2015, Sol Phys, 290, 1703

\bibitem[{Okamoto {et~al.}(2008)Okamoto, Tsuneta, Lites, Kubo, Yokoyama, Berger, Ichimoto, Katsukawa, Nagata, Shibata, {et~al.}}]{okamoto2008emergence}
Okamoto, T.~J., Tsuneta, S., Lites, B.~W., Kubo, M., Yokoyama, T., Berger, T.~E., Ichimoto, K., Katsukawa, Y., Nagata, S., Shibata, K., {et~al.} 2008, The Astrophysical Journal, 673, L215

\bibitem[{Ouyang {et~al.}(2015)Ouyang, Yang, \& Chen}]{ouyang2015flux}
Ouyang, Y., Yang, K., \& Chen, P. 2015, The Astrophysical Journal, 815, 72

\bibitem[{{Ouyang} {et~al.}(2017){Ouyang}, {Zhou}, {Chen}, \& {Fang}}]{Ouyang2017}
{Ouyang}, Y., {Zhou}, Y.~H., {Chen}, P.~F., \& {Fang}, C. 2017, \apj, 835, 94

\bibitem[{Pariat \& D{\'e}moulin(2012)}]{pariatEstimationSquashingDegree2012}
Pariat, E. \& D{\'e}moulin, P. 2012, A\&A, 541, A78

\bibitem[{Patsourakos {et~al.}(2020)Patsourakos, Vourlidas, T{\"o}r{\"o}k, Kliem, Antiochos, Archontis, Aulanier, Cheng, Chintzoglou, Georgoulis, {et~al.}}]{patsourakos2020decoding}
Patsourakos, S., Vourlidas, A., T{\"o}r{\"o}k, T., Kliem, B., Antiochos, S., Archontis, V., Aulanier, G., Cheng, X., Chintzoglou, G., Georgoulis, M., {et~al.} 2020, Space Science Reviews, 216, 1

\bibitem[{Perri {et~al.}(2022)Perri, Leitner, Brchnelova, Baratashvili, Ku{\'z}ma, Zhang, Lani, \& Poedts}]{perri2022coconut}
Perri, B., Leitner, P., Brchnelova, M., Baratashvili, T., Ku{\'z}ma, B., Zhang, F., Lani, A., \& Poedts, S. 2022, The Astrophysical Journal, 936, 19

\bibitem[{Porth {et~al.}(2014)Porth, Xia, Hendrix, Moschou, \& Keppens}]{porth2014mpi}
Porth, O., Xia, C., Hendrix, T., Moschou, S., \& Keppens, R. 2014, The Astrophysical Journal Supplement Series, 214, 4

\bibitem[{{Schatten} {et~al.}(1969){Schatten}, {Wilcox}, \& {Ness}}]{1969SoPh....6..442S}
{Schatten}, K.~H., {Wilcox}, J.~M., \& {Ness}, N.~F. 1969, \solphys, 6, 442

\bibitem[{{Scherrer} {et~al.}(2012){Scherrer}, {Schou}, {Bush}, {Kosovichev}, {Bogart}, {Hoeksema}, {Liu}, {Duvall}, {Zhao}, {Title}, {Schrijver}, {Tarbell}, \& {Tomczyk}}]{2012SoPhsdohmi}
{Scherrer}, P.~H., {Schou}, J., {Bush}, R.~I., {Kosovichev}, A.~G., {Bogart}, R.~S., {Hoeksema}, J.~T., {Liu}, Y., {Duvall}, T.~L., {Zhao}, J., {Title}, A.~M., {Schrijver}, C.~J., {Tarbell}, T.~D., \& {Tomczyk}, S. 2012, \solphys, 275, 207

\bibitem[{{Schou} {et~al.}(2012){Schou}, {Scherrer}, {Bush}, {Wachter}, {Couvidat}, {Rabello-Soares}, {Bogart}, {Hoeksema}, {Liu}, {Duvall}, {Akin}, {Allard}, {Miles}, {Rairden}, {Shine}, {Tarbell}, {Title}, {Wolfson}, {Elmore}, {Norton}, \& {Tomczyk}}]{2012SoPhsdohmiground}
{Schou}, J., {Scherrer}, P.~H., {Bush}, R.~I., {Wachter}, R., {Couvidat}, S., {Rabello-Soares}, M.~C., {Bogart}, R.~S., {Hoeksema}, J.~T., {Liu}, Y., {Duvall}, T.~L., {Akin}, D.~J., {Allard}, B.~A., {Miles}, J.~W., {Rairden}, R., {Shine}, R.~A., {Tarbell}, T.~D., {Title}, A.~M., {Wolfson}, C.~J., {Elmore}, D.~F., {Norton}, A.~A., \& {Tomczyk}, S. 2012, \solphys, 275, 229

\bibitem[{Scott {et~al.}(2017)Scott, Pontin, \& Hornig}]{scott2017magnetic}
Scott, R.~B., Pontin, D.~I., \& Hornig, G. 2017, The Astrophysical Journal, 848, 117

\bibitem[{{Song} {et~al.}(2017){Song}, {Chen}, {Li}, {Li}, {Zhao}, {He}, {Duan}, {Cheng}, \& {Zhang}}]{2017songfilament}
{Song}, H.~Q., {Chen}, Y., {Li}, B., {Li}, L.~P., {Zhao}, L., {He}, J.~S., {Duan}, D., {Cheng}, X., \& {Zhang}, J. 2017, \apjl, 836, L11

\bibitem[{Tassev \& Savcheva(2017)}]{tassevQSLSquasherFast2017a}
Tassev, S. \& Savcheva, A. 2017, ApJ, 840, 89

\bibitem[{Titov \& D{\'e}moulin(1999)}]{titov1999basic}
Titov, V. \& D{\'e}moulin, P. 1999, Astronomy and Astrophysics, 351, 707

\bibitem[{Titov {et~al.}(2014)Titov, T{\"o}r{\"o}k, Mikic, \& Linker}]{titov2014method}
Titov, V., T{\"o}r{\"o}k, T., Mikic, Z., \& Linker, J.~A. 2014, The Astrophysical Journal, 790, 163

\bibitem[{Titov(2007)}]{titovGeneralizedSquashingFactors2007}
Titov, V.~S. 2007, ApJ, 660, 863

\bibitem[{Titov {et~al.}(2018)Titov, Downs, Miki{\'c}, T{\"o}r{\"o}k, Linker, \& Caplan}]{titov2018regularized}
Titov, V.~S., Downs, C., Miki{\'c}, Z., T{\"o}r{\"o}k, T., Linker, J.~A., \& Caplan, R.~M. 2018, The Astrophysical Journal Letters, 852, L21

\bibitem[{Titov {et~al.}(2002)Titov, Hornig, \& D{\'e}moulin}]{titov2002theory}
Titov, V.~S., Hornig, G., \& D{\'e}moulin, P. 2002, Journal of geophysical research: Space physics, 107, SSH

\bibitem[{T{\"o}r{\"o}k {et~al.}(2010)T{\"o}r{\"o}k, Berger, \& Kliem}]{torokWritheHelicalStructures2010}
T{\"o}r{\"o}k, Berger, \& Kliem. 2010, A\&A, 516, A49

\bibitem[{{Turk} {et~al.}(2011){Turk}, {Smith}, {Oishi}, {Skory}, {Skillman}, {Abel}, \& {Norman}}]{turk2011pythonyt}
{Turk}, M.~J., {Smith}, B.~D., {Oishi}, J.~S., {Skory}, S., {Skillman}, S.~W., {Abel}, T., \& {Norman}, M.~L. 2011, \apjs, 192, 9

\bibitem[{van Ballegooijen \& Martens(1989)}]{van1989formation}
van Ballegooijen, A.~A. \& Martens, P. 1989, Astrophysical Journal, Part 1 (ISSN 0004-637X), vol. 343, Aug. 15, 1989, p. 971-984., 343, 971

\bibitem[{Verbeke {et~al.}(2022)Verbeke, Baratashvili, \& Poedts}]{verbekeICARUSNewInner2022}
Verbeke, C., Baratashvili, T., \& Poedts, S. 2022, A\&A, 662, A50

\bibitem[{{Wang} {et~al.}(2022){Wang}, {Chen}, {Ding}, \& {Lu}}]{2022ApJWangcanRMHDplasma}
{Wang}, C., {Chen}, F., {Ding}, M., \& {Lu}, Z. 2022, \apjl, 933, L29

\bibitem[{{Wang} {et~al.}(2023){Wang}, {Chen}, {Ding}, \& {Lu}}]{2023ApJWangcanRMHDforces}
---. 2023, \apj, 956, 106

\bibitem[{{Wang} \& {Muglach}(2007)}]{2007wangformation_filament}
{Wang}, Y.~M. \& {Muglach}, K. 2007, \apj, 666, 1284

\bibitem[{{Wang} \& {Sheeley}(1992)}]{1992ApJ...392..310W}
{Wang}, Y.~M. \& {Sheeley}, N.~R., J. 1992, \apj, 392, 310

\bibitem[{Warnecke \& Peter(2019)}]{warnecke2019data}
Warnecke, J. \& Peter, H. 2019, Astronomy \& Astrophysics, 624, L12

\bibitem[{Wheatland {et~al.}(2000)Wheatland, Sturrock, \& Roumeliotis}]{wheatland2000optimization}
Wheatland, M., Sturrock, P., \& Roumeliotis, G. 2000, The Astrophysical Journal, 540, 1150

\bibitem[{Wiegelmann {et~al.}(2006)Wiegelmann, Inhester, \& Sakurai}]{wiegelmann2006preprocessing}
Wiegelmann, T., Inhester, B., \& Sakurai, T. 2006, Solar Physics, 233, 215

\bibitem[{{Xia} \& {Keppens}(2016)}]{2016xia_ronyfilament}
{Xia}, C. \& {Keppens}, R. 2016, \apj, 823, 22

\bibitem[{Xia {et~al.}(2018)Xia, Teunissen, El~Mellah, Chan{\'e}, \& Keppens}]{xia2018mpi}
Xia, C., Teunissen, J., El~Mellah, I., Chan{\'e}, E., \& Keppens, R. 2018, The Astrophysical Journal Supplement Series, 234, 30

\bibitem[{Xu {et~al.}(2020)Xu, Zhu, \& Guo}]{xuThreedimensionalReconstructionFilament2020}
Xu, Y., Zhu, J., \& Guo, Y. 2020, ApJ, 892, 54

\bibitem[{{Yang} {et~al.}(2024){Yang}, {Xia}, \& {Guo}}]{https://doi.org/10.5281/zenodo.13831036}
{Yang}, K.~E., {Xia}, C., \& {Guo}, Y. 2024, Kai-E-Yang/QSL: QSL v1

\bibitem[{Zhao {et~al.}(2017)Zhao, Xia, Keppens, \& Gan}]{zhao2017formation}
Zhao, X., Xia, C., Keppens, R., \& Gan, W. 2017, The Astrophysical Journal, 841, 106

\bibitem[{{Zhong} {et~al.}(2021){Zhong}, {Guo}, \& {Ding}}]{2021NatCoZhongze}
{Zhong}, Z., {Guo}, Y., \& {Ding}, M.~D. 2021, Nature Communications, 12, 2734

\bibitem[{Zhou {et~al.}(2020)Zhou, Chen, Hong, \& Fang}]{zhou2020simulations}
Zhou, Y., Chen, P., Hong, J., \& Fang, C. 2020, Nature Astronomy, 4, 994

\end{thebibliography}

\end{document}